\documentstyle[prl,twocolumn,aps]{revtex}
\input{epsf}
\begin{document}
\draft
\twocolumn[\hsize\textwidth\columnwidth\hsize\csname @twocolumnfalse\endcsname
\title{Delocalization of two-particle ring near the Fermi level of 2d Anderson
model}

\author{J. Lages and D. L. Shepelyansky$^{*}$}

\address {Laboratoire de Physique Quantique, UMR 5626 du CNRS, 
Universit\'e Paul Sabatier, F-31062 Toulouse Cedex 4, France}

\date{February 18, 2000}

\maketitle

\begin{abstract}
We study analytically and numerically the problem of two particles 
with a long range attractive interaction on a two-dimensional (2d) 
lattice with disorder. It is shown that below some critical disorder
the interaction creates delocalized coupled states near the Fermi level.
These states appear inside well localized noninteracting phase and
have a form of two-particle ring which diffusively propagates over 
the lattice.
\end{abstract}
\pacs{PACS numbers:  72.15.Rn, 71.30+h, 74.20.-z, 05.45.Mt}
\vskip1pc]

\narrowtext


Recently a great deal of attention has been attracted to the problem of 
interaction effects in  disordered systems with Anderson localization
\cite{moriond,hamburg}. From the theoretical point of view the problem is
rather nontrivial. Indeed, even if a great progress has been reached in 
the theoretical investigation of the properties of localized eigenstates 
\cite{mirlin} still the analytical expressions for interaction matrix 
elements between localized states are lacking. In spite of these theoretical
difficulties it has been shown recently that a repulsive or attractive
interaction between particles can destroy localization and lead to a 
propagation of pairs in the noninteracting localized phase. This two 
interacting particles (TIP) effect has been studied recently by different
groups \cite{ds94,imry,pichard,vonoppen,song,frahm,ortuno,schreiber} and
it has been understood that the delocalization of TIP pairs is 
related to the enhancement of interaction in systems with complex,
chaotic eigenstates. Such an enhancement had been already known for
parity violation induced by the weak interaction in heavy nuclei
\cite{suflam} where the interaction is typically increased by a factor
of thousand. However, since there the two-body interaction is really weak
the final result still remains small. On the contrary, for TIP pairs
in the localized phase the enhancement of interaction qualitatively 
changes the dynamics leading to a coherent propagation of TIP on a 
distance $l_c$ being much larger than the pair size and 
one-particle localization length 
$l_1$. The enhancement factor $\kappa$ is determined by the density of
two-particle states $\rho_2$, coupled by interaction, 
and the interaction induced
transition rate $\Gamma_2$ between noninteracting eigenstates, so that
$\kappa = \Gamma_2 \rho_2$. At $\kappa \sim 1$ the interaction matrix element
becomes comparable with two-particle level spacing and the Anderson 
localization starts to be destroyed by interaction. For excited states
the TIP density $\rho_2$ is significantly larger than the one-particle
density $\rho$ and the delocalization can be reached for relatively weak
interaction if $l_1$ is large. However, when the excitation energy $\epsilon$
above the Fermi level decreases then $\rho_2$ becomes smaller and it
approaches the one-particle 
density $\rho$ at low energy: $\rho_2 \approx \epsilon \rho^2$ \cite{imry}.
As a result the value of $\kappa$ also drops with $\epsilon$ so that the
delocalization of TIP pairs practically disappears near the Fermi energy.
This result has been found in \cite{imry,jacquod} in the approximation of the
frozen Fermi sea created by fermions. Recent numerical studies of TIP pairs
with short range interaction near the Fermi level \cite{lages}  confirmed
these theoretical expectations.

\begin{figure}
\vskip -0.2cm
\epsfxsize=8.5cm
\epsfysize=10.5cm
\epsffile{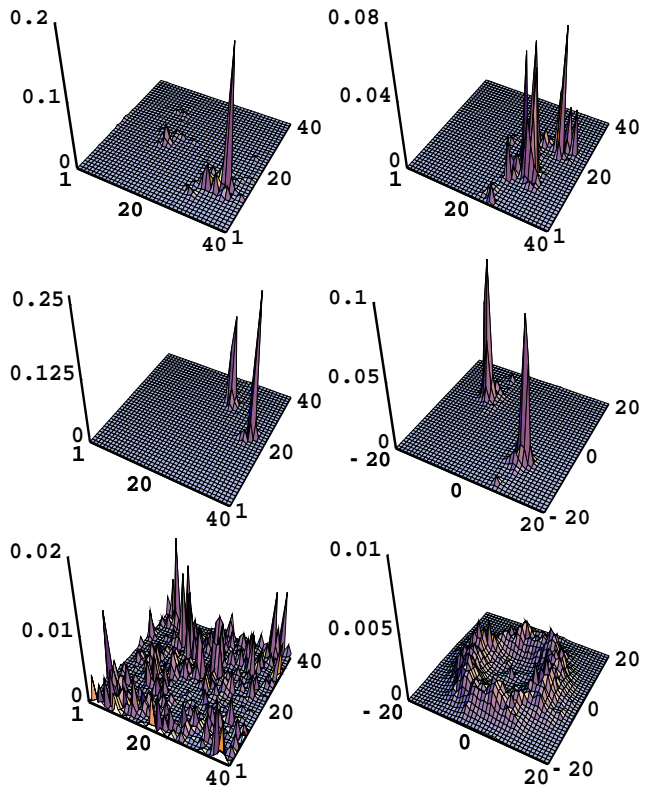}
\vglue 0.2cm
\caption{Probability distributions $f$ and $f_d$ for TIP in 2d disordered
lattice of size $L=40$, and interaction of radius $R=12$ and
width $\Delta R = 1$. Left column, one-particle probability $f$
for $W=8V$: ground state at $U=0$ (top); ground state with binding 
energy $\Delta E = -1.05 V$ at $U=-2V$ (middle); coupled state with
$\Delta E = -0.19 V$ at $U=-2V$ (bottom). Right column: 
$f$ for coupled state, compare to bottom left, at $W=12V$ and $U=-2V$ with 
$\Delta E \approx -0.19 V$(top); inter-particle
distance probability
$f_d$ related to the middle left case (middle); $f_d$
related to the bottom left case (bottom). All data are
shown for the same disorder realisation. Color 2d density plots for these data 
are given in Fig. 1bis of Appendix.
} 
\label{fig1}
\end{figure}

In this paper we discuss another type of situation in which TIP 
delocalization (see Fig. 1) takes place mainly due to geometrical reasons and
not due to the relation $\rho_2 \gg \rho$. As a result the TIP pair 
can be delocalized in a close vicinity to the Fermi level that opens
new interesting possibilities for interaction
induced delocalization in the localized noninteracting phase.
To study this new situation we investigate a model with a long range
attractive interaction between particles in the 2d Anderson model. 
In this case the particles can rotate around their center of mass,
being far from each other and 
keeping the same energy, while the center can move randomly in two dimensions.
As a result the system has effectively three degrees of freedom that makes
it rather similar to the case of one particle in
the 3d Anderson model where delocalization takes place at sufficiently
weak disorder. A somewhat similar situation has been studied recently
for particles with Coulomb interaction but only excited states
were considered there and the delocalization was attributed to the large ratio
$\rho_2 / \rho$ \cite{ds99}. Here we show that in fact the conditions for
delocalization are much less restrictive. 

To illustrate the above ideas let us first discuss the case of only two
particles with attractive interaction $U(r) < 0$ in the 2d 
Anderson model described by the Schr\"odinger equation
\begin{equation}
\label{ham}
\begin{array}{c}
(E_{\mathbf n_1}+E_{\mathbf n_2}+U({\mathbf n_1-n_2}))\psi_{\mathbf n_1,n_2}
+V(\psi_{\mathbf n_1+1,n_2}\\
+\psi_{\mathbf n_1-1,n_2}
+\psi_{\mathbf n_1,n_2+1}+\psi_{\mathbf n_1,n_2-1})=E\psi_{\mathbf n_1,n_2}.
\end{array}
\end{equation}
\noindent Here ${\mathbf n_{1,2}}$ are the indices of the two particles on the 
2d lattice with $L^2$ sites and periodic boundary conditions,
$V$ is the hopping
between nearby sites and the random on-site one-particle energies 
$E_{\mathbf n_{1,2}}$ are homogeneously distributed in the interval 
$[-W/2,W/2]$. The long range attractive interaction depends only on the
distance between particles $r=\|\mathbf{n_1-n_2}\|$ and is equal to a 
constant $U<0$ if $|r-R| \leq \Delta R$ and zero otherwise. The value of
$r$ is determined as the minimal inter-particle distance on the periodic 
lattice. Thus
the interaction takes place only inside a ring of radius $R$ and width
$\Delta R$, and we assume that $R \gg \Delta R \ge 1$. For $U=0$ the 
eigenstates are given by the product of two one-particle (noninteracting)
eigenstates which are always localized in 2d in a presence 
of disorder \cite{abrahams}.

In the limit of very strong attractive interaction $|U| \gg V$ the TIP 
coupled states form the energy band of width $\simeq 16V$ around 
$E \simeq -|U|$ (we consider only the states symmetric in respect to 
particle interchange). For states in this band the particles are located 
always inside the ring which center can move over the 2d lattice. Since
$|U| \gg V$ these states are decoupled from all other states
with particles outside the ring. In the ring the
Schr\"odinger equation is in fact rather similar to the case of
3d Anderson model of one particle. In this analogy the number of sites inside the ring 
$M_R \approx 2 \pi R \Delta R$ determines the effective number of 2d planes
placed  one over another
in the third $z$-dimension (length size $L_z = M_R$). In this 3d model the
effective strength of disorder is approximately 
$2 W$ since the diagonal term is now the sum of two
$E_{\mathbf n}$ values. Also one site is coupled with $Z=8$ neighbours 
contrary to $Z=6$ for 3d case (assuming $\Delta R \gg 1$). Since in 3d the
Anderson transition at the band center takes place at 
$W_c = 2.75 Z V = 16.5 V$ \cite{ohtsuki}, we expect that TIP states 
inside the ring 
will be delocalized in the middle of the band when $2 W / Z V = 2.75$ that 
gives the transition at $W_{c2} \approx 11 V$. This estimate is in
agreement with numerical simulations of the model ({\ref{ham})
\cite{caldara}. Of course, since the
size in the third direction is finite the eigenstates will be eventually
localized. But their localization length $l_c$ will make a sharp jump from
$l_c \sim 1$ at $W > W_{c2}$ to $l_c \sim \exp( g ) \gg 1$ at $W<W_{c2}$ 
that follows from the standard scaling theory in 2d
\cite{mirlin,abrahams,ds99}. Here $g$ is the 
conductance of the quasi-two-dimensional layer of width $L_z = M_R$. As usual
$g = E_c / \Delta_1$ where $E_c= D / L^2$ is the Thouless energy,
$\Delta_1 \sim V / (L^2 L_z)$ is the level spacing \cite{mirlin} and
the diffusion rate in the lattice model is $D \sim V(V / W)^2$. As a result 
for $W < W_{c2}$ the TIP delocalization length jumps to exponentially large
value $l_c \sim \exp(2 \pi R \Delta R (W_{c2} / W)^2 )$. In these estimates
we assumed that $l_1 > \Delta R > 1$ since if $\Delta R \gg l_1$ the majority
of states inside the ring are noninteracting and can be presented as the
product of one-particle eigenstates. We also note that for $W < W_{c2}$
there is an energy interval around the band center with delocalized states
where the TIP ring diffuses with the rate $D_2 \sim V (W_{c2}/V)^2$.
When $W$ decreases the mobility edge approaches the bottom of the band
as it happens in 3d Anderson model.

The above arguments presented for the case $|U| \gg V$ indicate that it is
possible to have a similar TIP delocalization at moderate value of $U \sim V$
near the Fermi level. To investigate this case we rewrite the equation
(\ref{ham}) in the basis of the noninteracting eigenstates that gives
\begin{eqnarray}
\label{tipeq}
(E_{m_1}+E_{m_2})\chi_{m_1, m_2} & + & 
U \sum_{{m^{'}_1}, {m^{'}_2}} Q_{m_1, m_2, {m^{'}_1}, {m^{'}_2}}
 \chi_{{m^{'}_1}, {m^{'}_2}} \nonumber \\
  & = & E\chi_{m_{1}, m_{2}}.
\end{eqnarray}
Here $\chi_{m_1, m_2}$ are eigenfunctions of the TIP problem written in the
basis of one-particle eigenstates $\phi_m$ with eigenenergies $E_m$. The
matrix $UQ_{m_1, m_2, {m^{'}_1}, {m^{'}_2}}$ represents the two-body matrix
elements of interaction $U({\mathbf n_1-n_2})$ between noninteracting 
eigenstates $|\phi_{m_1} \phi_{m_2}\rangle$ and 
$|\phi_{m^{'}_1} \phi_{m^{'}_2}\rangle$.
The Fermi sea is determined by the restriction of the summation in (\ref{tipeq})
to $m^{(')}_{1,2} > 0$ with energies $E_{m^{(')}_{1,2}} > E_F$, where $E_F$ 
is  the Fermi energy related to the filling factor $\mu$. We choose the case
with half filling $\mu = 1/2$ for which $E_F \approx 0$. In this way our model
corresponds to the approximation of frozen Fermi sea successfully used 
for the Cooper problem \cite{cooper}. As it was done by Cooper we also
introduce the high energy cut-off defined by the condition 
$1 \leq m^{'}_1 + m^{'}_2 \leq M$. This rule determines an effective phonon
frequency $\omega_D \propto M/L^2$. We fix $\alpha = L^2/M \approx 15$
since $\omega_D$ should be independent of the system size $L$ \cite{note1}.
We checked that the results are not affected by a variation of $\alpha$ in
few times. The first studies of the TIP model with frozen Fermi sea was
done by Imry \cite{imry} with the aim to take into account the effect
of finite fermionic density and then was also analyzed in \cite{jacquod}.
Recently a similar model was investigated for the case of Hubbard 
attraction in 3d \cite{lages}.
\begin{figure}
\epsfxsize=8cm
\epsfysize=8cm
\epsffile{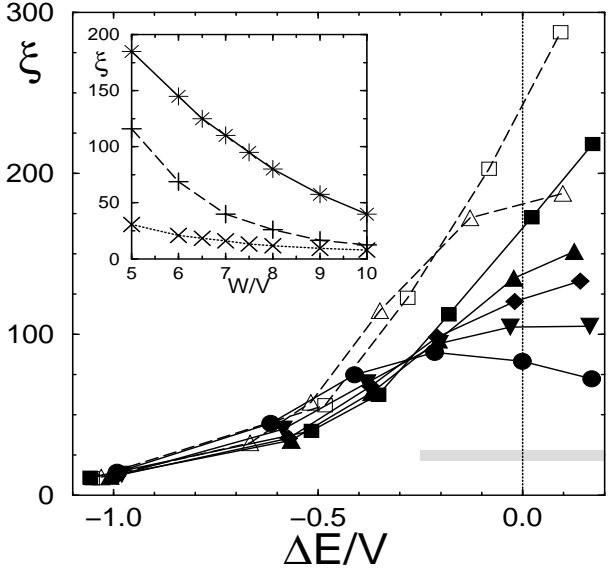}
\vglue 0.2cm
\caption{Dependence of IPR $\xi$ on the binding energy $\Delta E / V$ for
$U=-2V$, $W=8V$ and $\Delta R = 1$: $L=20$ $(\circ)$, $L=22$ 
(triangle down), $L=26$ (diamond), $L=30$ (triangle up), $L=40$ (square);
full/empty symbols are for $R=8 / R=12$; the shaded band shows the variation
of $\xi$ at $E_F$ and $U=0$ for $20 \leq L \leq 40$. Insert shows dependence
of $\xi$ on disorder $W/V$ at $R=8$ for the ground state at $U=-2V, L=30$
$(\times)$ and at $U=0, L=30$ $(+)$, and for the states at 
the delocalization border with binding 
energy $\Delta E \approx \Delta E_c$ at $U=-2V$ $(\ast)$.} 
\label{fig2}
\end{figure}

To study the eigenstate properties of our model we diagonalize
numerically the Hamiltonian (\ref{tipeq}) and rewrite the eigenfunctions
in the original lattice basis. In this way we determine the two-particle
probability distribution $F({\mathbf n_1, n_2})$ from which we extract 
the one particle probability 
$f({\mathbf n_1})= \sum_{\mathbf n_2} F({\mathbf n_1, n_2})$
and the probability of inter-particle distance 
$f_d({\mathbf r})=\sum_{\mathbf n_2} F({\mathbf r+n_2, n_2})$ with 
${\mathbf r= n_1 - n_2}$. The binding energy of an eigenstate in 
(\ref{tipeq}) is $\Delta E = E - 2 E_F \approx E$ since $E_F \approx 0$. 
For the ground state with energy $E_g$ the coupling energy is
$\Delta = 2E_F - E_g$.
The typical examples of probability distributions are shown in Fig. 1.
They clearly show that the ground state in the presence of interaction
remains localized and the particles stay on distance $R$ from each 
other. However, there are states with negative binding energy 
($\Delta E < 0$)
which are delocalized by interaction and for which the particles move
around the ring in agreement with discussion of model (\ref{ham})
at $|U| \gg V$. We stress that this delocalization of coupled
states ($\Delta E < 0$) takes place in the well localized one-particle
phase. However, at very strong disorder this delocalization disappears
(see top right case in Fig. 1). 

To analyze the delocalization of states with negative binding energy $\Delta E$
in a more 
quantitative way we determine the inverse participating ratio (IPR) $\xi$ for 
one-particle probability 
$1/\xi=\langle \sum_{{\mathbf n}}f^2({\mathbf n}) \rangle$, where brackets
mark the averaging over 100 disorder realisations. In this way $\xi$ gives
the number of lattice sites occupied by one particle in an eigenstate. 
The dependence of $\xi$ on $\Delta E$ and $W$ is shown in Fig. 2 for 
different lattice sizes $L$ in the presence of interaction. This figure 
shows that near the ground state the interaction creates states which
are even more localized than in the absence of interaction ($\xi$ is
significantly smaller than at $U=0$, see insert Fig. 2). 
\begin{figure}
\epsfxsize=8cm
\epsfysize=8cm
\epsffile{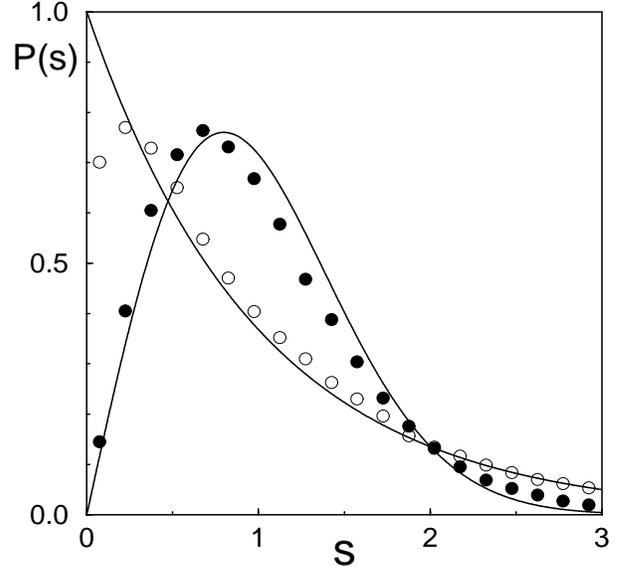}
\vglue 0.2cm
\caption{Level spacing statistics $P(s)$ for TIP coupled states at $L=30$,
$U=-2V$, $W=8V$, $R=8$, $\Delta R = 1$ in the localized phase near the 
ground state inside the energy interval $-\Delta< \Delta E < -3\Delta/4$
$(\circ)$ and in the delocalized phase for energies inside 
$-\Delta/4 < \Delta E < 0$ 
$(\bullet)$; here $\Delta E_c \approx -0.3 \Delta$ and 
the statistics is done over 3000 disorder realisations. Full lines show
the Poisson distribution and the Wigner surmise.}
\label{fig3}
\end{figure}
\noindent In fact for 
$-\Delta < \Delta E < \Delta E_c < 0$ the IPR value even slightly drops
with the increase of $L$. However for the states with binding energy
$\Delta E_c < \Delta E < 0$ the situation becomes different and $\xi$
grows significantly with $L$ while the change of IPR at $U=0$ with $L$
is rather weak (see shaded band in Fig. 2). 
The critical value of the binding energy 
$\Delta E_c$ can be defined as such an energy at which $\xi$ remains 
independent of $L$. In this way $\Delta E_c$ determines the mobility edge
for coupled states so that at given $U$ and $W$ the TIP eigenstates 
are localized for $-\Delta < \Delta E < \Delta E_c $ while for 
$\Delta E_c < \Delta E < 0$ the states becomes delocalized (see an 
example in Fig. 1). In agreement with this picture $\xi$ varies up to 
30 times when $\Delta E$ changes from $ -\Delta$ up to 
$0$. This variation grows with $L$ and 
the interaction radius $R$ since the system becomes 
more close to the effective 3d Anderson
model as it was discussed above. The qualitative change of the structure 
of the eigenstates leads also to a change in the level spacing statistics
$P(s)$ (Fig. 3). Near the ground state the statistics is close to the Poisson 
distribution $P_P(s)= \exp (-s)$ typical for the localized Anderson phase
\cite{shklovskii} while for $\Delta E_c < \Delta E < 0$ it approaches
to the Wigner surmise $P_W(s)= \pi s \exp(-\pi s^2 /4) \, /2$ corresponding
to the delocalized phase \cite{shklovskii}. 
\begin{figure}
\epsfxsize=8cm
\epsfysize=8cm
\epsffile{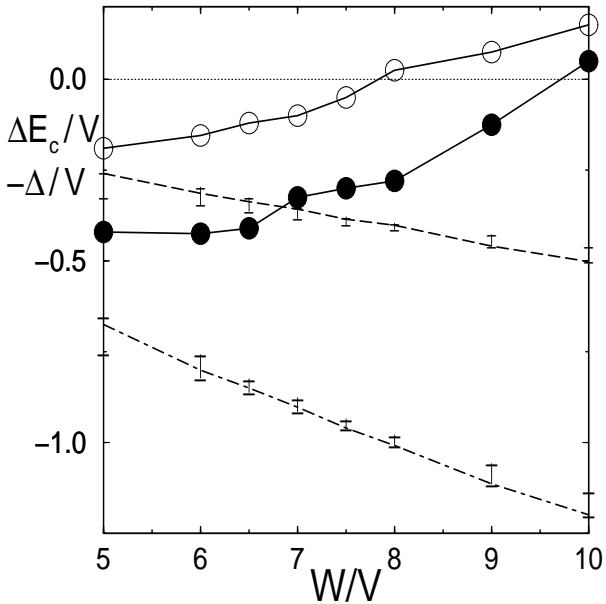}
\vglue 0.2cm
\caption{Dependence of the delocalization border $\Delta E_c$ and the 
binding energy of the ground state $-\Delta$ on disorder $W/V$. 
The values of $\Delta E_c$ are shown by $(\circ)$/$(\bullet)$ for 
$U=-V \, / \, -2V$; the values of $-\Delta$ are shown by upper/lower
dashed line for $U=-V \, / \, -2V$ and $L=30$ (vertical intervals give
variation for $20 \leq L \leq 30$).}
\label{fig4}
\end{figure}

\enlargethispage{\baselineskip}
The variation of the delocalization border $\Delta E_c$ for TIP coupled
states with disorder strength and interaction is shown in Fig. 4. While
the coupling energy $\Delta$ grows with $U$ and $W$, the mobility edge 
$\Delta E_c <0$, on the contrary, disappears at strong $W$.
According to the data of Fig. 4 all states with binding energy 
$\Delta E <0$ become localized for $W > W_{c2} \approx 9.5V$ $(U=-2V)$ 
and $W > W_{c2} \approx 8V$
$(U=-V)$. This shows that at weaker interaction a weaker disorder is required 
to have delocalized coupled states. As it follows from Fig. 4, at small 
disorder $W$ the delocalization border $\Delta E_c$ becomes closer and 
closer to the ground state. This means that at weak disorder the 
delocalization will take place for excited states with low energy.
For $W \ll W_{c2}$ and $U \sim -V$ the diffusion rate of delocalized TIP ring
can be estimated as $D_2 \sim V(W_{c2}/W)^2$ \cite{note2}. Further studies are
required to determine the dependence of $W_{c2}$ on $W$ at $|U| \ll V$.
 
In conclusion, our results show that long range attractive interaction
between two particles in 2d leads to the appearance of delocalized diffusive
states near the Fermi level inside the well localized noninteracting phase.
It would be interesting to understand what will be the consequences of this
delocalization for real many-body fermionic problem with attractive 
interaction. It is possible that obtained results will be also relevant
for electrons with Coulomb repulsion. Indeed, in this case at very weak 
disorder each electron oscillates near an equilibrium position and the 
two-body interaction can be considered as an effective harmonic attraction
\cite{ponomarev}.

We thank O.P.Sushkov for stimulating discussions,
and the IDRIS in Orsay and the CICT in Toulouse for access to 
their supercomputers.

\enlargethispage{2\baselineskip}

\newpage
\thispagestyle{empty}
\begin{minipage}[b]{\textwidth}
\vskip -1cm
\begin{center}
\Huge Appendix
\end{center}
\begin{figure}
\begin{center}
\epsfxsize=15.5cm
\epsfysize=22cm
\epsffile{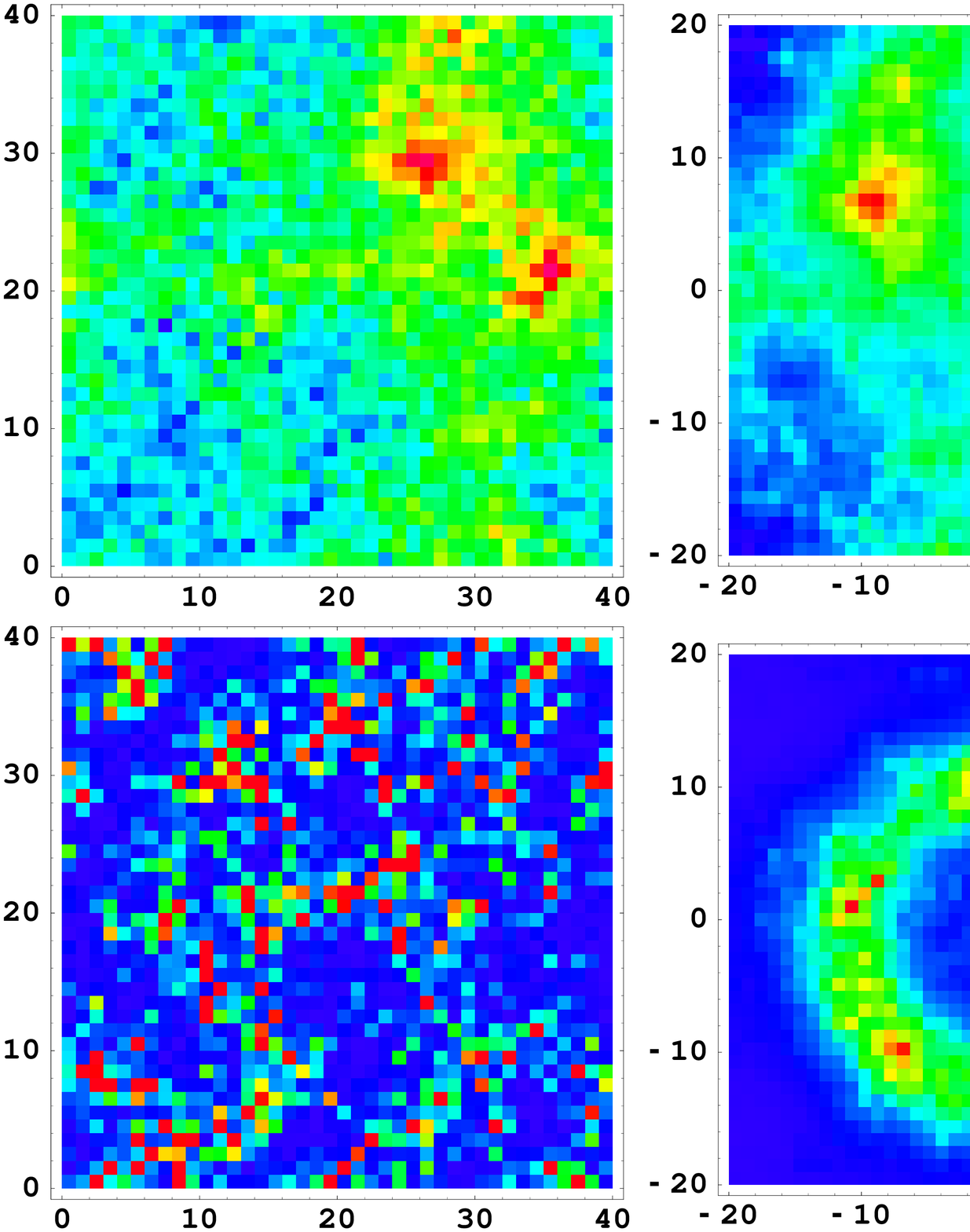}
\end{center}
\end{figure}
\vskip -1cm
\normalsize
\enlargethispage{3cm}
FIG. 1bis. Color 2d density plots for the data of Fig.1 with the 
same ordering of figures.
Blue corresponds to the minimum of the 
probabilty distribution and red to the maximum. The first four figures
are drawn in logarithmic scale while two figures at the bottom
are in linear scale.
Blue/Red color corresponds to: $f=1.3\times 10^{-10}/f=0.2$
(top left), $f=1.1\times 10^{-9}/f=0.26$ (middle left), 
$f=1.5\times 10^{-6}/f=0.014$ (bottom left), $f=1.14\times 10^{-11}/f=0.073$
(top right), $f_d=3.8\times 10^{-7}/f_d=0.1$ (middle right),
$f_d=1.5\times 10^{-5}/f_d=0.0032$ (bottom right). 
\end{minipage}


\begin{thebibliography}{99}
\bibitem{note1} In this case we found that $\omega_D/V \approx 0.055 W/V +
0.20$.
\bibitem{note2} We expect that $D_2$ will drop with $\Delta E$ growth
                from $\Delta E_c$ to $0$; indeed, for $\Delta E > 0$ the 
                particles are in average more separated from each other
                and their interaction becomes effectively smaller.
\end{thebibliography}
\end{document}